\documentclass[aps,prl,twocolumn,floatfix,superscriptaddress,amsmath,amssymb,longbibliography]{revtex4-2}
\usepackage{xcolor,bm,lineno,multirow,times}
\usepackage{physics}
\usepackage{rotating}
\usepackage{verbatim}
\usepackage{graphicx,tabularx}
\usepackage[sort&compress]{natbib}
\usepackage[T1]{fontenc}
\usepackage{soul}
\usepackage{dcolumn}
\usepackage{bm}
\usepackage[version=4]{mhchem}
\usepackage{booktabs}
\usepackage{gensymb}
\usepackage{color}
\usepackage{ulem}
\usepackage[colorlinks=true,urlcolor=blue,linkcolor=magenta,citecolor=magenta]{hyperref}

\newcommand{\knso}{K$_2$Ni(SeO$_3$)$_2$}

\allowdisplaybreaks

\begin{document}

\title{Spin dynamics of the spin-1 triangular lattice Heisenberg antiferromagnet K$_2$Ni(SeO$_3$)$_2$}

\author{Chaebin Kim}
\email{ckim706@gatech.edu}
\affiliation{School of Physics, Georgia Institute of Technology, Atlanta, Georgia 30332, USA}
\author{Sathvik Nallapati}
\affiliation{School of Physics, Georgia Institute of Technology, Atlanta, Georgia 30332, USA}
\author{E. A. Ghioldi}
\affiliation{Department of Physics and Astronomy, The University of Tennessee, Knoxville, Tennessee 37996, USA}
\author{Long Chen}
\affiliation{Department of Physics and Astronomy, The University of Tennessee, Knoxville, Tennessee 37996, USA}
\author{Alexander I. Kolesnikov}
\affiliation{Neutron Scattering Division, Oak Ridge National Laboratory, Oak Ridge, Tennessee 37831, USA}
\author{Haidong Zhou}
\affiliation{Department of Physics and Astronomy, The University of Tennessee, Knoxville, Tennessee 37996, USA}
\author{Shang-Shun Zhang}
\affiliation{Department of Physics and Astronomy, The University of Tennessee, Knoxville, Tennessee 37996, USA}
\author{Cristian D. Batista}
\affiliation{Department of Physics and Astronomy, The University of Tennessee, Knoxville, Tennessee 37996, USA}
\affiliation{Quantum Condensed Matter Division and Shull-Wollan Center, Oak Ridge National Laboratory, Oak Ridge, Tennessee 37831, USA}
\author{Martin Mourigal}
\email{mourigal@gatech.edu}
\affiliation{School of Physics, Georgia Institute of Technology, Atlanta, Georgia 30332, USA}

\date{\today}

\begin{abstract}
Strong quantum fluctuations and unconventional spin dynamics are well established in the spin-1/2 triangular lattice Heisenberg antiferromagnet. However, their survival in the spin-1 case remains an open question. We investigate the spin dynamics of K$_2$Ni(SeO$_3$)$_2$, a nearly ideal spin-1 triangular lattice Heisenberg antiferromagnet, using inelastic neutron scattering. Below the ordering temperature $T_{\rm N}$, we observe coherent one-magnon excitations coexisting with a broad high-energy continuum. Two complementary approaches, a spectrally consistent $1/S$-corrected spin wave theory and a beyond-mean-field Schwinger boson theory, reproduce different facets of the continuum. Neither alone is complete, demonstrating substantial quantum fluctuations survive for $S\!=\!1$ and are reflected primarily in the spectral distribution of the continuum. Above $T_{\rm N}$, the continuum bandwidth is conserved while spectral weight is redistributed as magnons lose spatial coherence. Our results establish K$_2$Ni(SeO$_3$)$_2$ as a model triangular antiferromagnet, identifying bandwidth conservation and the distribution of spectral weight within the continuum as organizing principles to understand the spin dynamics of ordered quantum magnets beyond spin-1/2. Our results highlight the need for controlled calculations of the interacting multi-magnon sector of 2D antiferromagnets. 
\end{abstract}
\maketitle

\begin{figure*}[htb!]
    \centering
    \includegraphics[width=1.0\linewidth]{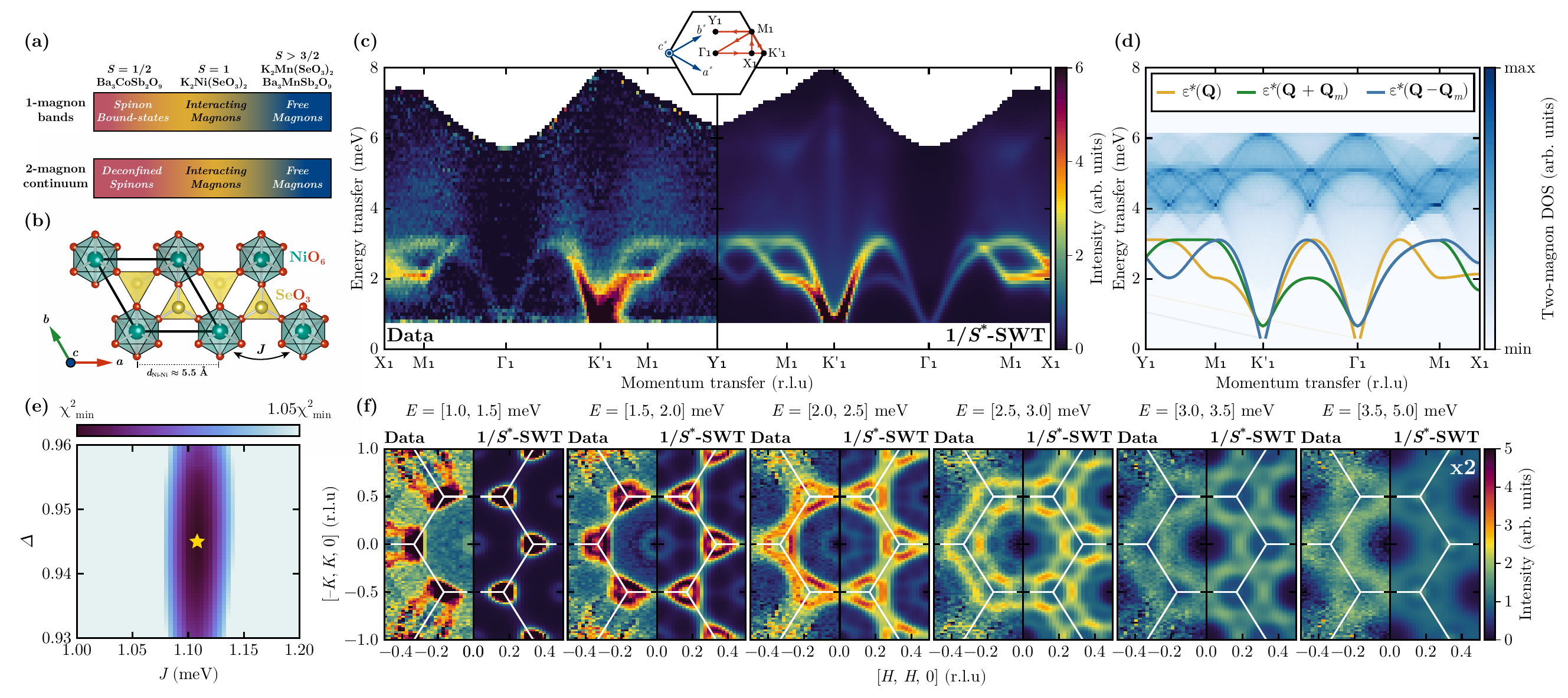}
        \caption{(a) Proposed theoretical treatment of 1-magnon and 2-magnon excitations in TLHAFs as a function of spin-$S$. (b) Crystal structure of $\mathrm{K_2Ni(SeO_3)_2}$, showing the layered triangular network of NiO$_6$ octahedrons with Ni$^{2+}$ ions connected by SeO$_3$ tetrahedrons. (c) The left panel shows momentum ($\bf Q$) and energy ($E$) transfer dependence of the inelastic neutron scattering intensity $I({\bf Q},E)$ measured at $T=2$~K along high-symmetry directions in the triangular-lattice Brillouin zone ${\bf Q} = (H,K,0)$ with out-of-plane integration $\Delta L= [-2, 2]$ r.l.u and $\Delta{\bf Q}_\perp =0.05$~\AA$^{-1}$. The right panel is the corresponding neutron scattering intensity calculated using spectrally-consistent $1/S^\ast$-SWT for optimized exchange parameters, convolved with an $E$-dependent Gaussian profile to match the dominant energy-resolution effects of the time-of-flight spectrometer. The faint vertical strike around the K point is a spurious signal from the off-shell calculation. (d) Comparison between the single-magnon dispersion at ${\bf Q}$ and its branches at  ${\bf Q} \pm {\bf Q}_{\rm m}$ umklapp-shifted by the propagation vector ${\bf Q}_{\rm m}$ of the underlying 120$^\circ$ magnetic order, together with the calculated two-magnon density of states. (e) Goodness-of-fit landscape obtained from a global least-squares fit of the $1/S^\ast$-SWT calculations to the data, for a two parameter model with nearest-neighbor antiferromagnetic exchange $J$ and easy-plane exchange anisotropy $\Delta$; the star marks the optimal parameter set used throughout this work. (f) Constant-$E$ slices of the scattering intensity in the $(H,K,0)$ plane for several energy windows from $1/S^\ast$-SWT (right panels) and experimental data (left panels) same $L$-integration as (c). The intensity of the last panel in (f) is scaled by a factor two.}
    \label{fig:1}
\end{figure*}

Heisenberg antiferromagnets (HAFMs) are among the simplest forms of quantum matter, providing stringent benchmarks for theory and numerics. Spin‑wave theory (SWT)~\cite{Oguchi1960}, which exploits broken symmetries to construct weakly interacting, low‑density magnon quasiparticles, often suffices to capture the behavior of ordered two‑ and three‑dimensional spin systems. However, for $S\!=\!1/2$ moments on even the most elementary quasi-two‑dimensional (2D) lattices, such as the nearest-neighbor square~\cite{Reger1988,  Manousakis1991}, triangular~\cite{Capriotti1999, Bernu1994}, and honeycomb~\cite{OitmaaHamer1992} HAFMs, quantum fluctuations~\cite{Ishikawa1975} suppress static order to $\lesssim 60$~\% of the classical value $\langle S^z \rangle\!=\!S$, even at low temperature. Such (relative) proximity to a quantum-melting point (QMP) accentuates what famously holds for any antiferromagnet~\cite{Anderson1984, ZhitomirskyRMP2013}, that from the point of view of the ordered state, \textit{``the spin of the spin wave ceases to exist''}, and challenges the large‑$S$ picture of a low density of non‑interacting magnons. Decades of effort have clarified the role of multi‑magnon continua~\cite{Canali1992}, which absorb spectral weight lost by static order, renormalize magnon energies, and enable spontaneous decay~\cite{Chernyshev2006, Chernyshev2009,Verresen2019}. While magnons \textit{are} undoubtedly the simplest description of low‑energy excitations of the low‑energy sector in ordered 2D HAFMs, a faithful account of the high‑energy, continuum excitations is not entirely settled for $S\!=\!1/2$, even for model compounds~\cite{Ma2016,Ito2017,Macdougal2020,Zhang2022BCAO,Sala2023YbCl3,Park2024BLCTO,Hernandez2025YbBr3,DallaPiazza2015,Elson2026}. Recent work has invoked either proximity to magnon fractionalization, by which magnon must be described as a two-spinon bound state~\cite{DallaPiazza2015, Shao2017PRX, Ghioldi2018, Zhang2019, Ghioldi2022}, or non‑perturbative corrections to composite‑magnon spectra~\cite{PowalskiUhrigSchmidt2015PRL, PowalskiSchmidtUhrig2018, Elson2026} to explain the distribution of spectral weight above the top of the magnon bands [Fig.~\ref{fig:1}(a)].
 
In triangular lattice materials, additional couplings and magnetic fields can tune proximity to a QMP~\cite{Kamiya2018,Xie2023} or suppress magnetic order entirely~\cite{Bordelon2019_NaYbO2_NatPhys,Scheie2024}. The most fundamental control of quantum fluctuations, however, is the spin length $S$ itself. Increasing $S$ enlarges the local Hilbert space (e.g., $S^z=0,\pm1$ for $S\!=\!1$) and often introduces on‑site anisotropies from the crystal environment. This, in turn, enables access to \textit{local} multi-magnon excitations~\cite{Bai2021} and their exotic dynamics~\cite{Sheng2025_Na2BaNiPO4_NatMater,Huang2025_pair_condensate_arXiv}. Yet the behavior of \textit{non-local} multi-quasiparticle spectra at the nearest‑neighbor Heisenberg point remains unsettled, partly because isostructural compounds with varying $S$ but comparable Hamiltonian complexity are rare. Two complementary material families, expanding from the model systems Ba$_3$CoSb$_2$O$_9$~\cite{Shirata2012,Susuki2013,Ma2016} and K$_2$Co(SeO$_3$)$_2$~\cite{Zhong2020_PRM}, now address this gap. The Ba‑based triangular perovskites span weakly anisotropic $S\!=\!1/2$ to $5/2$ and include extreme‑2D analogs~\cite{Rawl2017_Ba8CoNb6O24_PRB,Rawl2019_Ba8MnNb6O24_PRMat,Lu2018_A3NiNb2O9_PRB,Fak2017_Ba3NiSb2O9_PRB,Shu2023_Ba3MnSb2O9_PRB}, though obtaining large‑volume single crystals is often challenging. By contrast, the selenite series K$_2\, M$(SeO$_3$)$_2$ ($M$ = Co, Ni, Mn) is accessible in single‑crystal form and continuously tunes from a strongly anisotropic $S\!=\!1/2$ XXZ regime with supersolid physics for Co~\cite{Zhu2025_K2CoSeO3_npjQM,Chen2025} to a large‑$S$ near‑Heisenberg case for Mn~\cite{Zhu2026_K2MnSeO3_arXiv}.

In this Letter, we investigate the spin dynamics of the recently synthesized $S\!=\!1$ triangular‑lattice antiferromagnet K$_2$Ni(SeO$_3$)$_2$~\cite{Li2023} using inelastic neutron scattering. By mapping and modeling spin-spin correlations in the entire momentum ${\bf Q}$ and energy ${E}$ space, we confirm the nearly-ideal Heisenberg and nearest-neighbor properties of the compound, with a weak easy‑plane exchange anisotropy. We observe sharp low-energy magnon modes coexisting with a broad high-energy continuum extending to nearly twice the one‑magnon bandwidth. By comparing the data with spectrally-consistent non-linear SWT~\cite{ChernyshevZhitomirsky2006_PRL, ChernyshevZhitomirsky2009_PRB, Mourigal2013, Maksimov2016, Elson2026} and a beyond-mean-field Schwinger-Boson theory (BMF-SBT)~\cite{Ghioldi2018,Zhang2019,Zhang2022,Ghioldi2022}, we isolate which features are best explained in terms of multi-magnon processes versus putative fractionalized excitations. Our work shows that K$_2$Ni(SeO$_3$)$_2$ is the missing link between K$_2$Mn(SeO$_3$)$_2$ and Ba$_3$CoSb$_2$O$_9$. Taken together, these three systems~\cite{Ma2016,Ito2017,Macdougal2020,Zhu2026_K2MnSeO3_arXiv} experimentally demonstrate how quantum fluctuations impact spin dynamics as a function of spin-$S$ in the simplest model of quantum magnetism, the TLHAF. Our study highlights that an accurate description of high-energy continuum excitations in antiferromagnets provides a stringent benchmark not only for identifying novel states of matter, but also for diagnosing proximity to such phases.

\paragraph{Sample Properties.} Single-crystal specimens of K$_2$Ni(SeO$_3$)$_2$ were synthesized following Ref.~\cite{Li2023}. This coumpound crystallizes in the $\rm R\bar{3}m$ space-group with well-separated NiO$_6$ octahedra connected by SeO$_3$ trigonal pyramids, forming a triangular lattice in the $ab$-plane with $d_{\rm Ni-Ni} \approx 5.5$~\AA [Fig.~\ref{fig:1}(b)]. Our laboratory X-ray diffraction analysis on crushed single crystals shows no structural deviation within resolution down to $T\!=\!12$~K [See Supplemental Material, Fig.~\ref{sfig:xrd}]. The large nearest-neighbor distance yields Weiss temperatures of $\theta_{\rm W}^{ab(c)}\!=\!-72.4(-71.8)$~K~\cite{Li2023}, indicating moderately strong antiferromagnetic interactions and weak easy-plane anisotropy. The anisotropy is also reflected in isothermal magnetization [SM Fig.~\ref{sfig:overviewthermo}] and the ESR $g$‑factors ($g_{ab}\!=\!2.13>g_c\!=\!2.07$) for the Ni$^{2+}$ (3d$^8$) magnetic moments~\cite{Li2023}. A magnetic long-range ordering transition is evident at $T_{\rm N} = 7.1$~K with concomitant signatures in both specific heat ($\partial C/\partial T|_{T_{\rm N}} = 0$) and susceptibility ($\partial^2\chi/\partial^2 T|_{T_{\rm N}} = 0$) measurements [SM Fig.~\ref{sfig:fieldeplines}]. Under an applied magnetic field in the $ab$-plane, the peak in $C(T)$ splits above $\mu_0H\approx 9$~T, while the transition remains a single peak up to $14$~T with the field along $\boldsymbol{c}$. This behavior is consistent with weak easy-plane anisotropy and resembles Ba$_3$CoSb$_2$O$_9$~\cite{Susuki2013}. 

\begin{figure*}[th]
        \centering
        \includegraphics[width=1.0\linewidth]{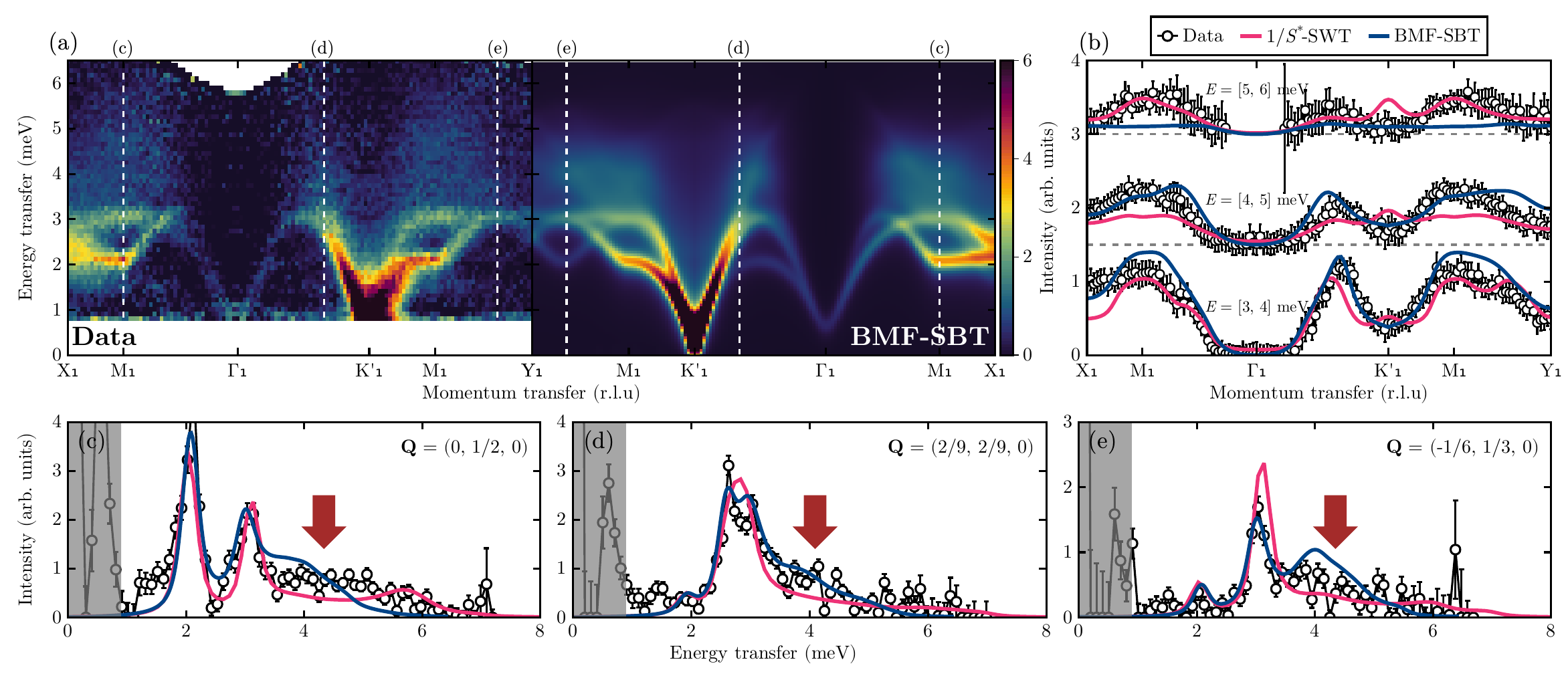}
        \caption{(a) Comparison between the $I({\bf Q},E)$ from Fig.~\ref{fig:1} and the corresponding $I_{\rm SBT}({\bf Q},E)$ from beyond-mean-field Schwinger boson theory (BMF-SBT) calculation. Vertical white dashed lines mark the momentum positions of the cuts presented in panels (c)–(e). (b) Momentum-dependent intensity profiles obtained by integrating the scattering intensity over the indicated energy windows. Experimental data (symbols) are compared with calculations from $1/S^\ast$-SWT (magneta lines) and BMF-SBT (blue lines). (c)–(e) Constant-momentum cuts as indicated in each panel. Open symbols represent the experimental data, and solid lines show the corresponding $1/S^\ast$-SWT and BMF-SBT calculations. The gray shaded regions are dominated by elastic and background contributions from the experiments. The red arrows mark the energy range where anomalous spectral weight is observed.}
        \label{fig:2}
\end{figure*}

\paragraph{Inelastic Neutron Scattering.} We conducted inelastic neutron scattering using the SEQUOIA time-of-flight spectrometer at the Spallation Neutron Source (SNS), Oak Ridge National Laboratory (ORNL)~\cite{Granroth2010}. Several dozens of hexagonal crystals, $m_{\rm tot} \approx$ 0.9 g, were co-aligned by x-ray Laue back-scattering, glued on an aluminum holder with hydrogen-free adhesive for a mosaic $\leq$ {$6\degree$}, and inserted in a liquid helium cryostat reaching $T\!=\!2$~K. The spectrometer operated with incident neutron energies $E_i\!=\!8$ and $11.5$~meV (elastic full-width-half-maximum resolutions of $\delta E\!=\!{0.18}$ and ${0.25}$~meV, respectively). The crystal assembly was aligned with the $ab$-plane horizontal and rotated about a vertical axis in steps of {1} degrees. The measured cross-section events were reduced to the magnetic scattering intensity $I({\bf Q},E)$ using {\scshape mantid}~\cite{MANTID}, symmetrized according to the $\bar{3}m$ Laue class, and projected using {\scshape shiver} algorithms~\cite{SHIVER} as two-dimensional slices. We adopt the hexagonal setting ${\bf Q} = H \boldsymbol{a}^\ast + K \boldsymbol{b}^\ast + L \boldsymbol{c}^\ast = (H,K,L)$ in reciprocal lattice units (r.l.u.) with high‑symmetry points corresponding to ${\rm \Gamma}=(0, 0, 0)$, ${\rm K}=(1/3,1/3,0)$, and ${\rm M}=(0,1/2,0)$.

Two-dimensional cuts through $I({\bf Q},E)$ at $T=2$~K are shown in Fig.~\ref{fig:1}(c,f), with the high-symmetry directions of the BZ defined in the inset. Coherent, gapless spin-wave excitations emerge from the ${\rm K}$ points of the BZ with a bandwidth of around 3 meV and a saddle point at the M points, but no roton-like minimum. The presence of three one-magnon branches is characteristic of the Heisenberg TLHAF 120$^\circ$ spin structure, and is associated with the propagation vector ${\bf Q}_{\rm m}=(1/3, 1/3, 0)$~\cite{Mourigal2013,Ma2016}. This ordering is confirmed by the development of sharp magnetic Bragg peaks below $T_{\rm N}$ at ${\bf Q} = (1/3,1/3,0)$ and $(2/3,2/3,0)$ [SM Fig.~\ref{sfig:elastic}]. The elastic scattering (integration over $\Delta E\!=\!\pm1.0$~meV) displays rods of scattering in the $L$-direction centered around $L=0$ and $L=\pm3$. Given the thermodynamic signatures, we interpret this diffuse scattering as resulting from $E$-integration in the absence of sizable dispersion along $L$. This evidences the quasi-two-dimensionality of the system also seen from the $L$-dependence of the zone-boundary excitations [SM Fig.~\ref{sfig:outofplane}]. In addition, we observe an intense, broad continuum-like signal above the one-magnon branches, reaching up to 6.5 meV (twice the one-magnon bandwidth), and associate it with 2-magnon excitations. This continuum is stronger than recently observed K$_2$Mn(SeO$_3$)$_2$~\cite{Zhu2026_K2MnSeO3_arXiv} but lacks the anomalous bandwidth observed for Ba$_3$CoSb$_2$O$_9$~\cite{Ito2017}. 

\paragraph{Model.} To describe the sharp spin excitations and continuum-like signal, we use the XXZ Hamiltonian: $\mathcal{H} = J\sum_{\langle i,j \rangle}[S_i^xS_j^x+S_i^yS_j^y+\Delta S_i^zS_j^z]$, where $J$ is the nearest-neighbor exchange interaction and $\Delta$ is the XXZ-exchange anisotropy ($\Delta = 0$ and $\infty$ correspond to the XY and Ising limits, respectively). A single-ion term, $D(S^z)^2$, is symmetry-allowed, but given the nearly octahedral orbital-singlet environment, its magnitude is second order in spin-orbit coupling and thus likely modest. Indeed, the apparent gapless spectrum sets $D$ below our resolution. Thus, we start with large-$S$ spin-wave theory (SWT) including $1/S$-corrections~\cite{Chernyshev2006,Mourigal2013,Maksimov2016}, necessary given that non-collinear spin order yields cubic interaction vertices. The magnon energy renormalization governed by the magnon self-energy $\Sigma({\bf q},\omega)$ reflects the hybridization of one- and two-magnon excitations. Provided kinematic conditions are met, spontaneous magnon decay occurs in some part of the BZ, but these conditions are quickly suppressed everywhere in the BZ in zero-field when $\Delta\leq\Delta_c=0.92$~\cite{Chernyshev2006,Maksimov2016}. 

As our focus is on comparing the bandwidth and the intensity of the low-energy and high-energy excitations, we need to account for the transfer of spectral weight between one- and two-magnon states, which are calculated at different perturbative order in a formal $1/S$-expansion. Instead, we calculate the diagonal components of the dynamical spin-structure factor (DSSF) $\mathcal{S}^{\alpha\alpha}({\bf Q},E)$ using a modified, ``spectrally-consistent'' off-shell version of Ref.~\cite{Mourigal2013}, recently developed for Ref.~\cite{Elson2026} [See SM Sec.~\ref{sec:si:NLSWT} for details]. In this approach, which we call $1/S^\ast$-SWT, the one-magnon dispersion is the solution of Dyson’s equation $\varepsilon_{{\bf Q}}^\ast$, and two-magnon states are constructed from it. Components of the DSSF are calculated in the local frame in terms of transverse and longitudinal components with respect to the ordered moments. In the transverse terms, the one-magnon spectral function allows for continuum sidebands; in the longitudinal channel, the two-magnon energies are renormalized but the magnons remain non-interacting. 

The result is shown in Fig.~\ref{fig:1}(c,f) after converting the DSSF to $I_{\rm SWT}({\bf Q},E)$ using the experiment’s polarization factor, form factor, energy-resolution, and momentum-integration ranges. We obtain $J\!=\!1.10(1)$~meV and $\Delta\!=\!0.94(2)$ by minimizing the $\chi^2$ between the data in Fig.~\ref{fig:1}(f) and $1/S^\ast$-SWT [See SM, Sec.~\ref{sec:si:chi2} for details]. The calculation agrees well with all aspects of the one-magnon spectra, including the ${\rm M}$-point known for its sensitivity to magnon interactions~\cite{Verresen2019} and the overall bandwidth of the continuum. While magnon decay is allowed in some parts of the BZ, the proximity to $\Delta_c$ and $S\!=\!1$ makes the decay rate negligible in this system. However, the $1/S^\ast$-SWT calculation misses the intensity distribution within the continuum and in particular its onset which starts just above the one-magnon band in the data, but it is pushed to higher energies in SWT. This can be traced to the density-of-states of two-renormalized-magnon [Fig.~\ref{fig:1}(d)] which concentrates between $E\!=\!4$ to $6$~meV.  While the sharp upper-edge of the continuum is an artifact that can be alleviated by invoking higher magnon-numbers, reproducing the continuum onset is a known limitation of SWT, even when transfer of spectral weight is allowed within the transverse channels~\cite{DallaPiazza2015,Elson2026}.

By treating the two‑magnon sector as renormalized yet non‑interacting, $1/S^\ast$-SWT shifts the two‑magnon density of states to too high energies and underestimates the low‑lying continuum intensity of the experiment. Similar early‑onset continua and spectral‑weight redistribution have been reported for the square-lattice zone-boundary anomaly, prompting competing interpretations in terms of magnon fractionalization \cite{DallaPiazza2015} versus non‑perturbative attractive magnon interactions, the latter currently providing the leading explanation \cite{PowalskiUhrigSchmidt2015PRL,PowalskiSchmidtUhrig2018}. For TLHAFs tuned near a QMP, such as Ba$_3$CoSb$_2$O$_9$ and KYbSe$_2$, spinon‑based descriptions successfully capture broad high‑energy continua while retaining sharp low‑energy magnons \cite{Ghioldi2022,Scheie2024}. Guided by these precedents and our data, we extend beyond–mean‑field Schwinger‑boson theory (BMF‑SBT) to the $S=1$ case [See SM Sec.~\ref{sec:si:SBT} for details]. In this framework, magnons appear as spinon bound states and the adjacent continuum as two‑spinon excitations within a single response function. Including Gaussian fluctuations of the auxiliary gauge fields is necessary to restore physical magnon poles and eliminates the unphysical single‑spinon poles present at the saddle point level.

In Fig.~\ref{fig:2}, we compare the INS data with the DSSF computed using BMF‑SBT and converted to the neutron scattering intensity \(I_{\rm SBT}(\mathbf{Q},E)\) following the same procedure as for SWT. The BMF‑SBT calculation yields a strong continuum that onsets immediately above the one‑magnon dispersion, consistent with the data, but it underestimates the continuum bandwidth. Constant‑energy cuts through the high‑energy signal along high‑symmetry paths in the BZ [Fig.~\ref{fig:2}(b)] expose complementary limitations of $1/S^\ast$–SWT and BMF‑SBT. The two calculations begin to differ at $E\!\approx\!4$~meV: in the $E\!=\![4,5]$~meV window BMF‑SBT reproduces the continuum intensity whereas $1/S^\ast$–SWT does not, whereas in the $E\!=\![5,6]$~meV window the situation reverses. The same trend appears in constant‑momentum cuts [Fig.~\ref{fig:2}(c–e)], where the continuum‑like signal is strongest and BMF‑SBT shows better agreement near $E\!\approx\!4$~meV. This can be traced back to the source of the continuum in each of the calculations. In the case of $1/S^\ast$-SWT, the high-energy continuum is dominated by longitudinal components with little contributions from transverse sidebands. However, BMF-SBT mainly produces the continuum in the transverse channel as the fractionalization of the one-magnon peak [See SM Sec.~\ref{sec:si:NLSWTvsSBT} for detailed channel-by-channel comparison between SBT and SWT].

\begin{figure}[b]
        \centering
        \includegraphics[width=1.0\linewidth]{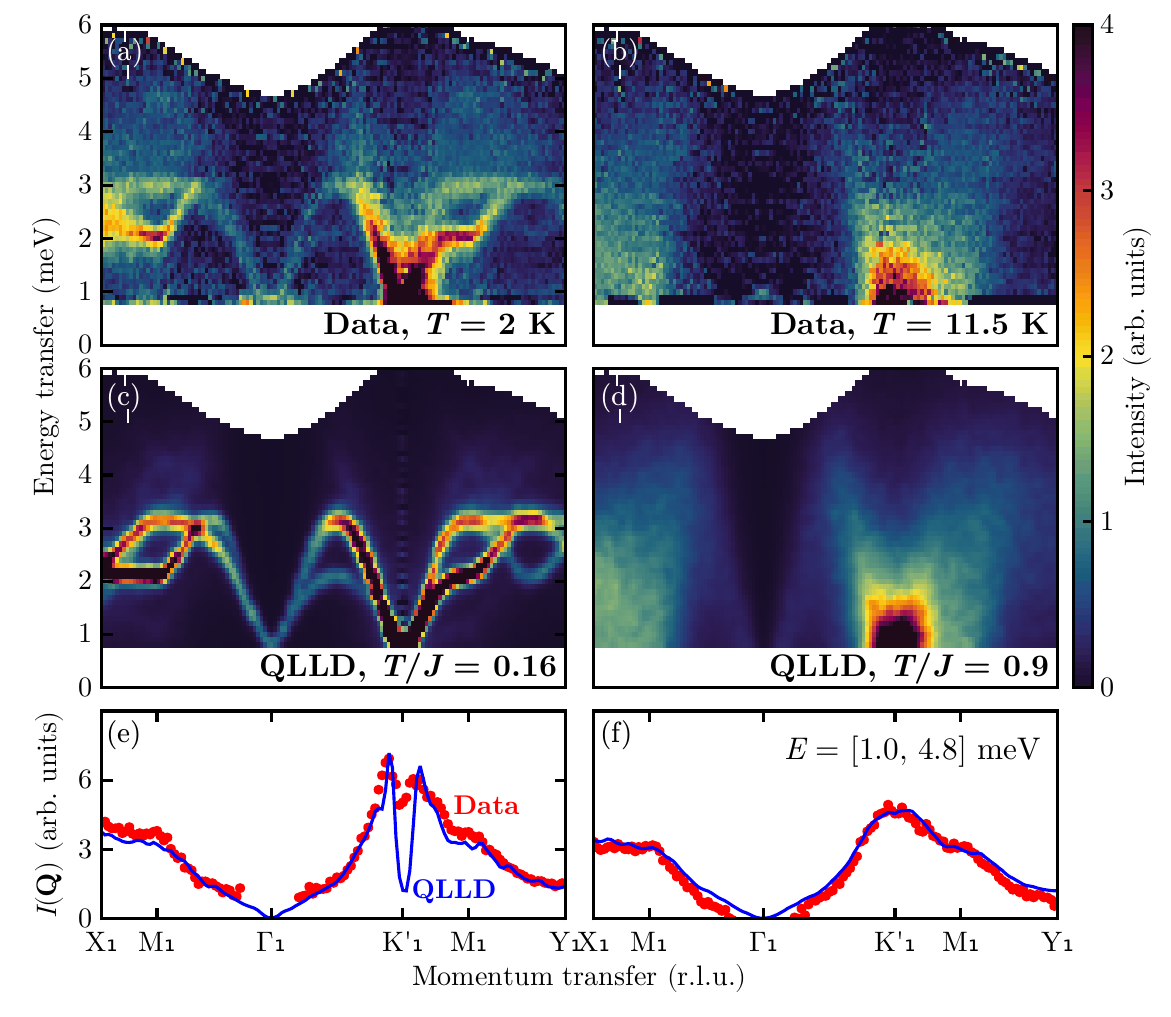}
        \caption{(a-b) Inelastic neutron scattering data for $E_i = 8$~meV at $T\!=\!2\ \mathrm{K}<\!T_{\rm N}$ and $T\!=\!11.5\ \mathrm{K}\!>\!T_{\rm N}$, respectively. (c-d) Corresponding QLLD calculation at the temperature scaled by the exchange interaction $J$. (e-f) The constant-energy cut with integration along the energy $E\!=\![1.0, 4.8]$~meV. Red dots are integrated neutron scattering data, and blue lines are integrated LLD simulations. }
        \label{fig:3}
\end{figure}

\paragraph{Discussion.} The comparison between the INS spectrum of a paradigmatic spin-one TLHAF and two quasiparticle-based calculations highlights the dual nature of the continuum excitations, with each approach capturing complementary aspects of the underlying physics. The two-spinon continuum of the SBT better describes the continuum above the single-magnon branch, suggesting that the spinon basis is more appropriate than the magnon basis in this energy regime. From the perspective of $1/S$-SWT, higher-order corrections beyond ${\cal O}(1/S)$ are required to account for the observed continuum features, but it’s unclear how to capture magnon interactions within the longitudinal two-magnon channel. In contrast, $1/S$-SWT accurately captures the continuum near the upper spectral edge, implying that the renormalized but essentially free magnon basis is more effective than the partially confined spinon basis. However, the present SBT calculation does not include four-spinon contributions, which may account for part of the remaining discrepancy near the upper spectral edge. This dichotomy is consistent with the entirely SWT-like continuum reported for $S=5/2$~\cite{Zhu2026_K2MnSeO3_arXiv} and recent conclusions for $S=1/2$ systems: a faithful account of the high-energy continuum in even ``elementary’’ Heisenberg antiferromagnets remains an open problem.

Having established the $T\!=\!0$ spin dynamics of K$_2$Ni(SeO$_3$)$_2$, we now extend our study to finite temperature to track how the excitations bandwidth and distribution of spectral‑weight redistributes across $T_N$. Figure~\ref{fig:3} summarizes this evolution by tracking the neutron scattering intensity across $T_{\rm N}\!=\!7.1$\,K. Below $T_N$, a continuum‑like signal persists up to $\approx\!6$\,meV alongside the one‑magnon mode. Strikingly, just above $T_N$ the continuum remains while the one-magnon poles broaden and the three modes associated with magnetic ordering merge. To model this behavior, we performed quantum‑corrected Landau–Lifshitz dynamics (QLLD) to calculate the DSSF and thus $I_{\rm LLD}({\bf Q},E)$ at finite temperature. Quantitative agreement with the paramagnetic data is achieved only when the quantum sum-rule is enforced, which effectively rescales the spin length by a factor $\kappa(T)$ and thereby increases the excitation bandwidth [See SM Sec.~\ref{sec:si:QLLD} for details and more temperature comparisons]. This result implies that the overall bandwidth of the multi‑magnon spectrum is approximately conserved into the correlated paramagnetic regime, even as magnons loose spatial coherence. An analogous conservation of continuum bandwidth is familiar from a 1D system, where the spinon continua at low temperature morph into a paramagnetic continuum at high temperature~\cite{Kim2025Q2C}.

K$_2$Ni(SeO$_3$)$_2$ is a model $S\!=\!1$ TLHAF with a weak easy-plane exchange anisotropy. In the magnetically ordered phase ($T\!<\!T_{\rm N}$), we resolve a high-energy continuum whose onset and upper edge are captured by complementary descriptions. At finite temperature, the continuum bandwidth is conserved across $T_{\rm N}$, with spectral weight redistributed as magnons lose spatial coherence. K$_2$Ni(SeO$_3$)$_2$, and its recently studied fraternal twin system K$_2$Mn(SeO$_3$)$_2$~\cite{Zhu2026_K2MnSeO3_arXiv}, are clean model materials that highlight the role of quantum fluctuations in the organization of spectral weight within the continuum of the TLHAF as a function of spin length $S$. These results establish two important principles for quasi-2D Heisenberg antiferromagnets beyond $S\!=\!1/2$: the continuum’s robustness to spin length and temperature, and the difficulty to describe its spectral lineshape. Our work calls to obtain a controlled, non-perturbative theory of the interacting multi-magnon sector of ordered magnets valid for $S \geq 1$ and finite $T$, capturing both the continuum onset and its upper edge.

\paragraph{Acknowledgements.} This work (C.K., S.N., S.S.Z., C.D.B., M.M.) was supported by the US Department of Energy, Office of Science, Basic Energy Sciences, Materials Sciences and Engineering Division under Award No. DE-SC0018660. The crystal growth at UTK (L.C. and H.D.Z.) was supported by the Air Force Office of Scientific Research under Grant No. FA9550-23-1-0502. Part of this research was conducted at the Spallation Neutron Source, a DOE Office of Science User Facility operated by the Oak Ridge National Laboratory. The beam time was allocated to SEQUOIA on proposal number IPTS-34576.  Part of this work (X-ray Diffraction) was performed in part at the Georgia Tech Institute for Matter and Systems, a member of the National Nanotechnology Coordinated Infrastructure (NNCI), which is supported by the National Science Foundation (Grant ECCS-2025462).

\paragraph{Data Availability.} The data that support the findings of this article will be made openly available.
\bibliographystyle{apsrev4-2} 
\bibliography{KNSO.bib}

\clearpage

\clearpage
\pagebreak
\widetext
\setcounter{section}{0}\setcounter{equation}{0}\setcounter{figure}{0}\setcounter{table}{0}
\renewcommand{\thesection}{S\arabic{section}}
\renewcommand{\theequation}{S\arabic{equation}}
\renewcommand{\thefigure}{S\arabic{figure}}
\renewcommand{\thetable}{S\arabic{table}}
\makeatletter
\setcounter{secnumdepth}{1}
\def\@seccntformat#1{\csname the#1\endcsname\quad}
\makeatother

\begin{center}
\textbf{\large \scshape Supplementary Information}
\end{center}

\section{X-ray diffraction}
\label{sec:si:xrd}

\begin{figure*}[h!]
        \centering
        \includegraphics[width=0.8\linewidth]{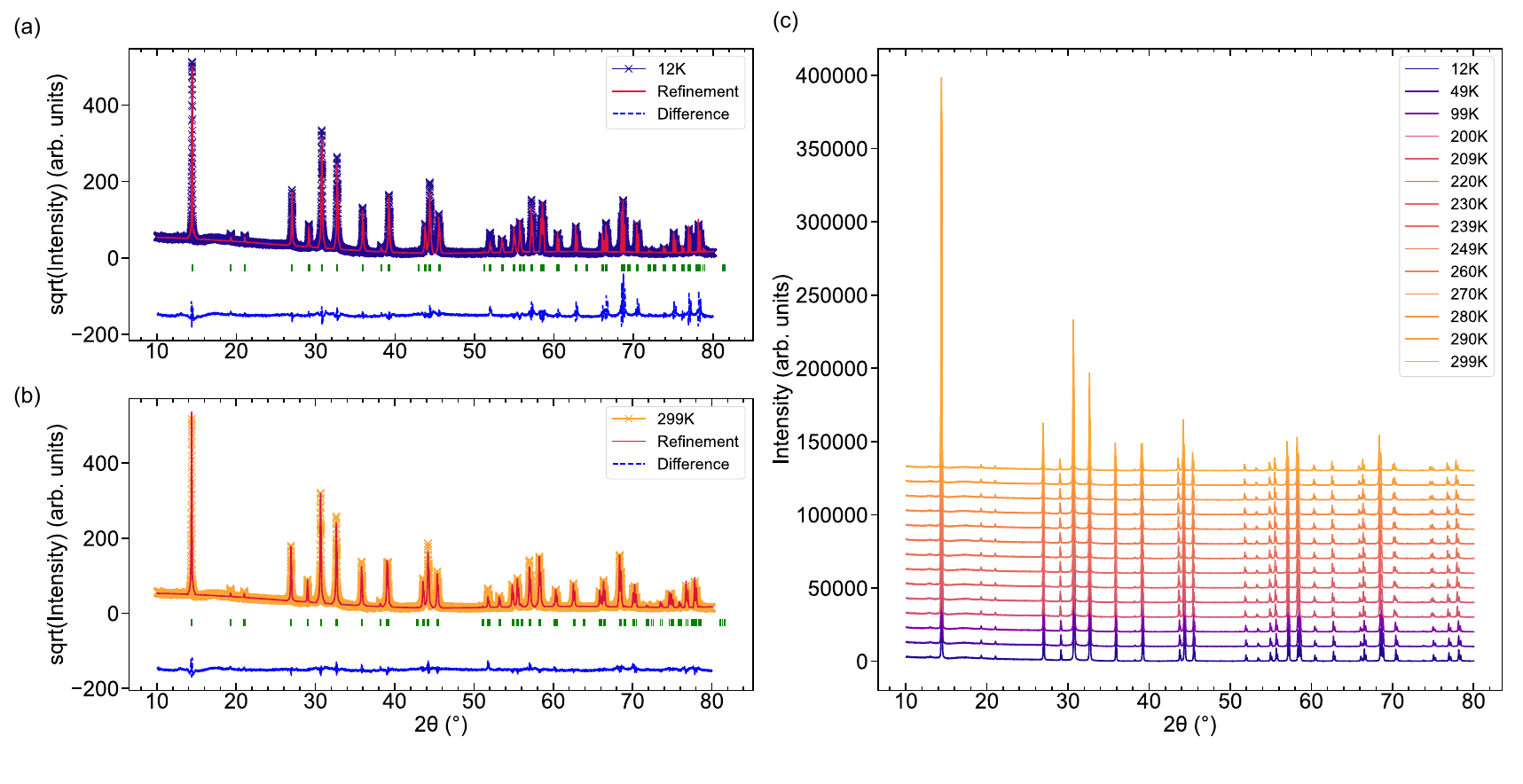}
        \caption{Powder X-ray diffraction data with best-fit structural Rietveld refinements for \knso at (a) $T=12$~K and (b) $T=299$~K. Nuclear peak positions are indicated by green vertical tick marks, best fit refinements are indicated by the solid red line, data is indicated by the colored circles, and the difference between the data and the best fit is indicated by the dashed blue line at the bottom of the panel. (c) Waterfall plot of all the data collected between $T=12$~K and $T=299$~K, showing no apparent structural phase transition.}
        \label{sfig:xrd}
\end{figure*}

Powder X-ray diffraction (PXRD) data were collected on a Rigaku SE X-Ray Diffractometer with Cu $K\alpha$ radiation ($ \lambda = 1.54056$ \r{A}) fitted with an Oxford Cryosystems Phenix cryostat. Multiple single crystals were ground into a fine powder, which was then spread evenly across a low-temperature cryogenic sample holder with Dupont Molykote cryogenic grease for thermal contact. This allowed for the collection of multiple data sets at different temperatures between $T=12$~K [Fig.~\ref{sfig:xrd}(a)] and 299~K [Fig.~\ref{sfig:xrd}(b)] to probe for any structural phase transitions, see Fig.~\ref{sfig:xrd}(c) for the full temperature dependence. Rietveld refinement of the PXRD data was performed using Fullprof~\cite{FullProf}. Refinements at $T=12$~K and $T=300$~K showed that the sample was phase pure, and remained in the ${\rm R\bar{3}m}$ space group down to $T = 12$~K. The lattice and structural parameters extracted from the refinements along with their goodness of fit are reported in Tab.~\ref{stab:xrd}.

\begin{table}[h!]
    \centering
    \begin{ruledtabular}
        \begin{tabular}{c c c c c c c}
            $T$ (K) & $\chi^2$ & $R_{wp}$ (\%) & a (\r{A}) & c (\r{A}) & $G_1$   & $G_2$   \\ \hline
            299     & 39.6     & 19.2       & 5.484989    & 18.463961  & 0.64981 & 0.11844 \\ 
            12      & 102      & 29.8       & 5.476611    & 18.378349  & 0.60518 & 0.35081 
        \end{tabular}
    \end{ruledtabular}
    \caption{Values of Structural parameters determined by Rietveld refinement against cryogenic Powder X-ray diffraction data of \knso. The parameters a and c are the cell length, $\chi^2$ and $R_{wp}$ measure goodness of fit, $G_1$ is a measure of preferred orientation, and values under 1 indicate that the ground crystals prefer to orient themselves in the plane, and $G_2$ is the fraction of the sample that is not textured.}
    \label{stab:xrd}
\end{table}

\section{Thermomagnetic Properties}
\label{sec:si:thermo}
\begin{figure*}[h!]
        \centering
        \includegraphics[width=0.8\linewidth]{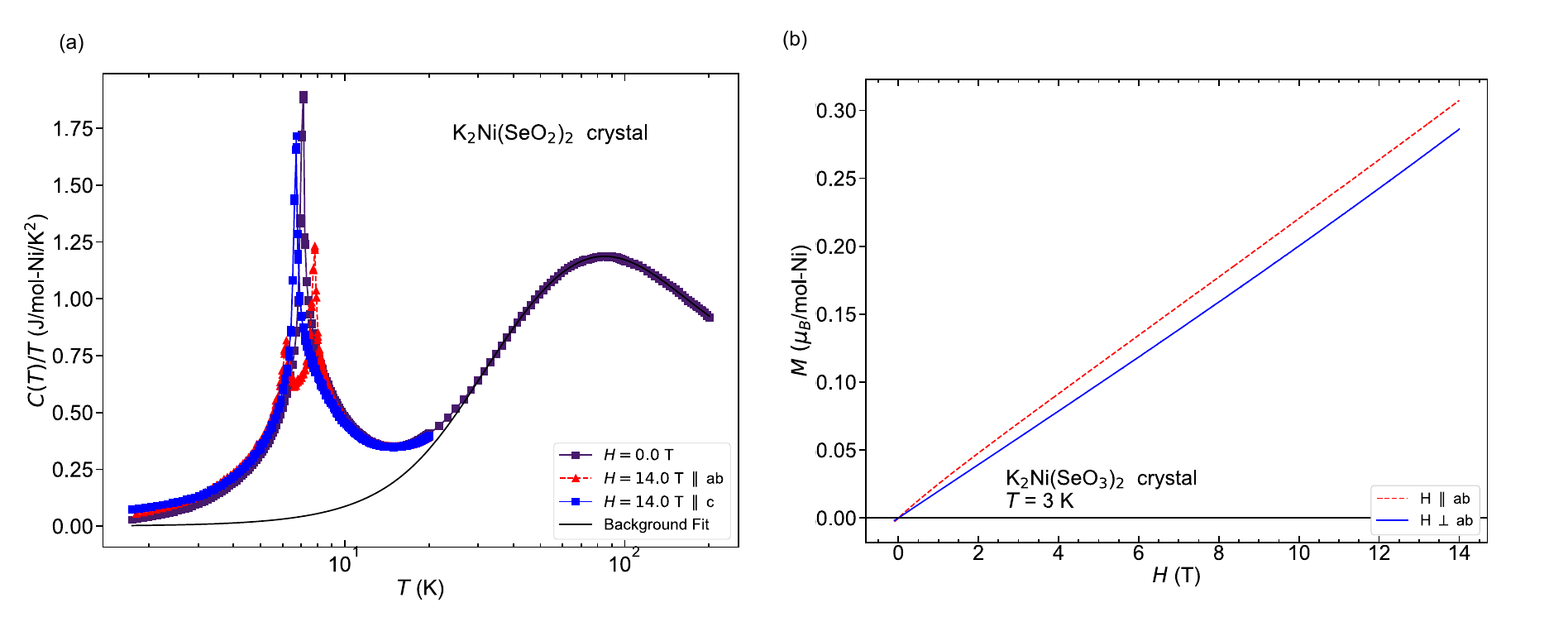}
        \caption{(a) Overview of specific heat $C(T)/T$ data for \knso taken up to 300~K at 0 and 14 T applied along different directions of the crystal. The solid line is the 0 T fit to the lattice contributions (b) Overview of isothermal magnetization $M(H)$ data with the external field being applied both in the plane of the crystal (red dashed line), and out of the plane (solid blue line). }
        \label{sfig:overviewthermo}
\end{figure*}

\begin{figure*}[h!]
        \centering
        \includegraphics[width=0.8\linewidth]{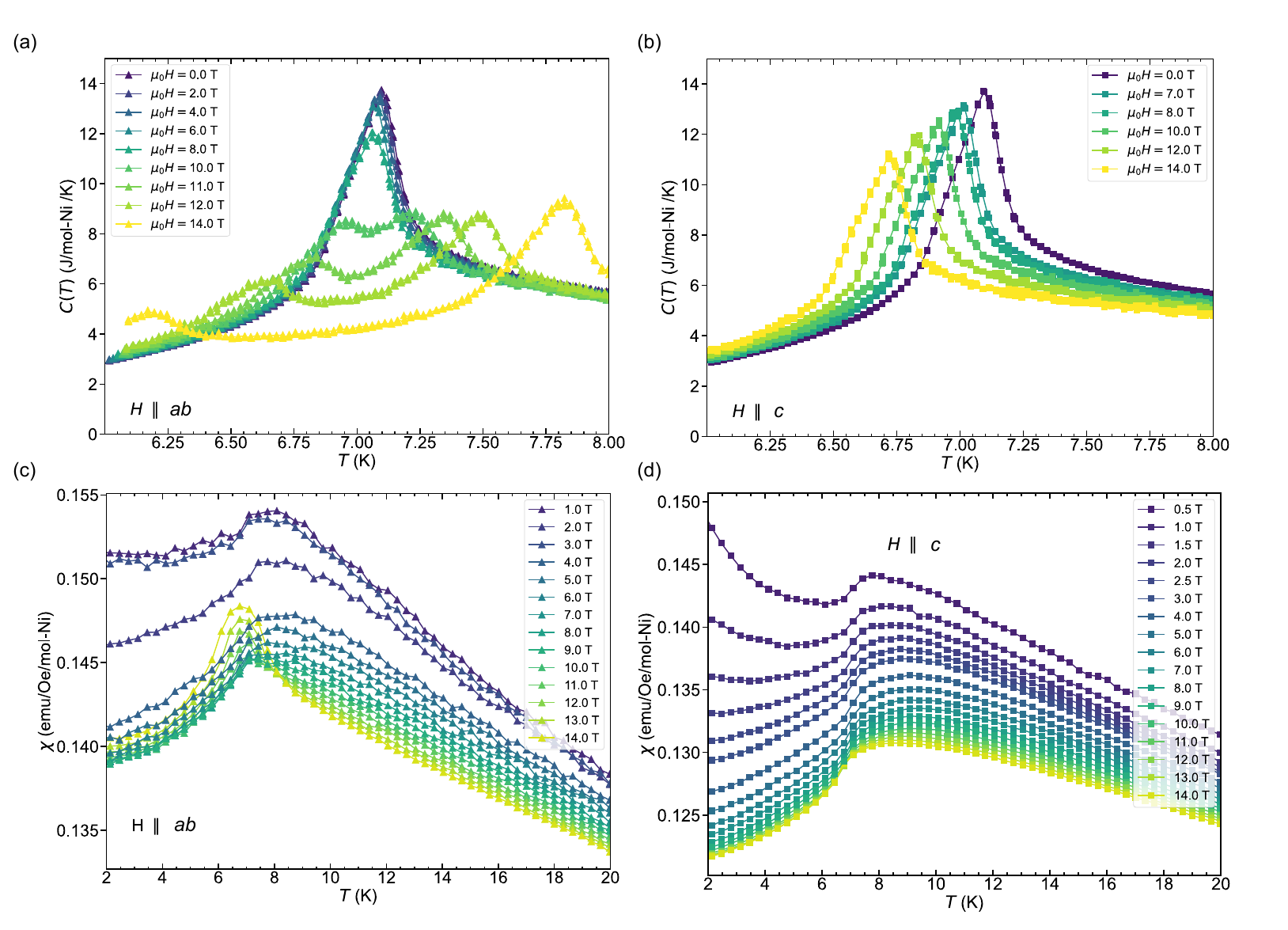}
        \caption{ (a) $C(T)$ data for \knso with an external magnetic field applied in the plane of the crystal. (b) $C(T)$ data for \knso with an external magnetic field applied out of the plane of the crystal. (c) $\chi(T)$ data for \knso with an external magnetic field applied in the plane of the crystal. (d) $\chi(T)$ data for \knso with an external magnetic field applied out of the plane of the crystal. }
        \label{sfig:fieldeplines}
\end{figure*}

Heat capacity $C(T)$ was measured using a Quantum Design PPMS for two field orientations, using 2–300 K data to fit and subtract a two-Debye plus one-Einstein lattice background. The sharp magnetic transition splits only for in-plane field while out-of-plane remains single as illustrated in Figs.~\ref{sfig:overviewthermo}(a) and \ref{sfig:fieldeplines}(a,b). Field-dependent $M(H)$ curves collected between 0 and 14 T using a Quantum Design PPMS at $T=3$~K  to reveal clear orientation-dependent anisotropy, see Fig.~\ref{sfig:overviewthermo}(b). Magnetization $M(T)$ was measured on an oriented single crystal from $T=2$ to $20$~K under $\mu_0H=0.5$–$14$ T with field applied both in and out of plane; the resulting susceptibilities $\chi(T)\!=\!M(T)/H$ are shown in Fig.~\ref{sfig:fieldeplines}(c,d).
\clearpage
\clearpage

\section{Elastic Scattering}
\begin{figure}[h!]
        \centering
        \includegraphics[width=0.4\linewidth]{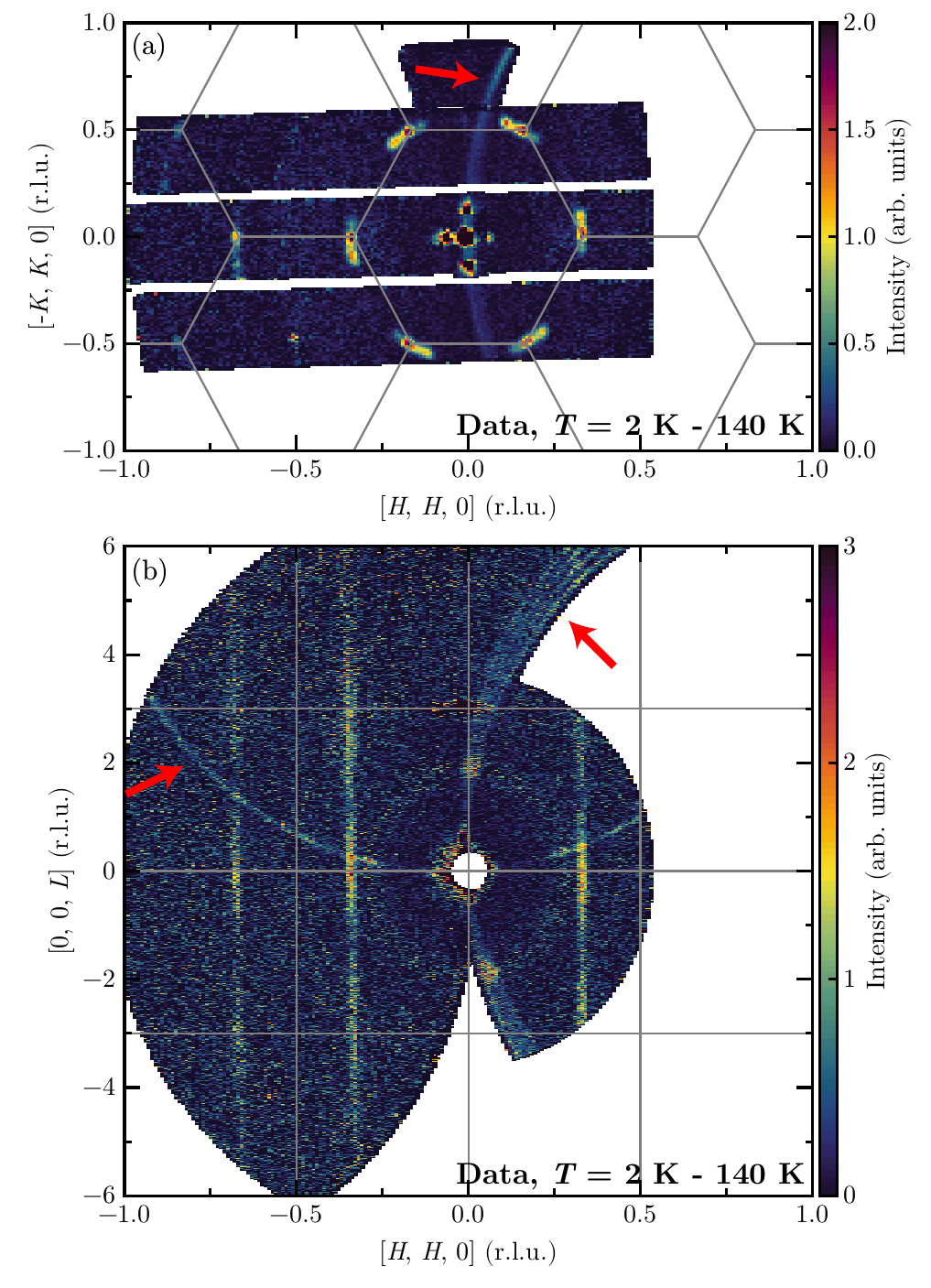}
        \caption{ (a) Elastic scattering in the $(H, K, 0)$-plane with the $E_i$ = 8 meV data at $T$ = 2 K integrated over $E=[-0.5,0.5]$~meV and $L=[-1,1]$~r.l.u. . The signal at the zone center corresponds to nuclear Bragg peaks and background originating from the $L$-integration. (b) Elastic scattering in the $(H, H, L)$ integrated over $E=[-0.5,0.5]$~meV and $K=[-0.05,0.05]$~r.l.u. . The data is not symmetrized and subtracted by $T$ = 140 K data. The red arrows indicate the artifact from the integration effect of the detector. }
        \label{sfig:elastic}
\end{figure}

\section{Out-of-plane direction spin-wave dispersion}
\begin{figure}[h!]
        \centering
        \includegraphics[width=0.5\linewidth]{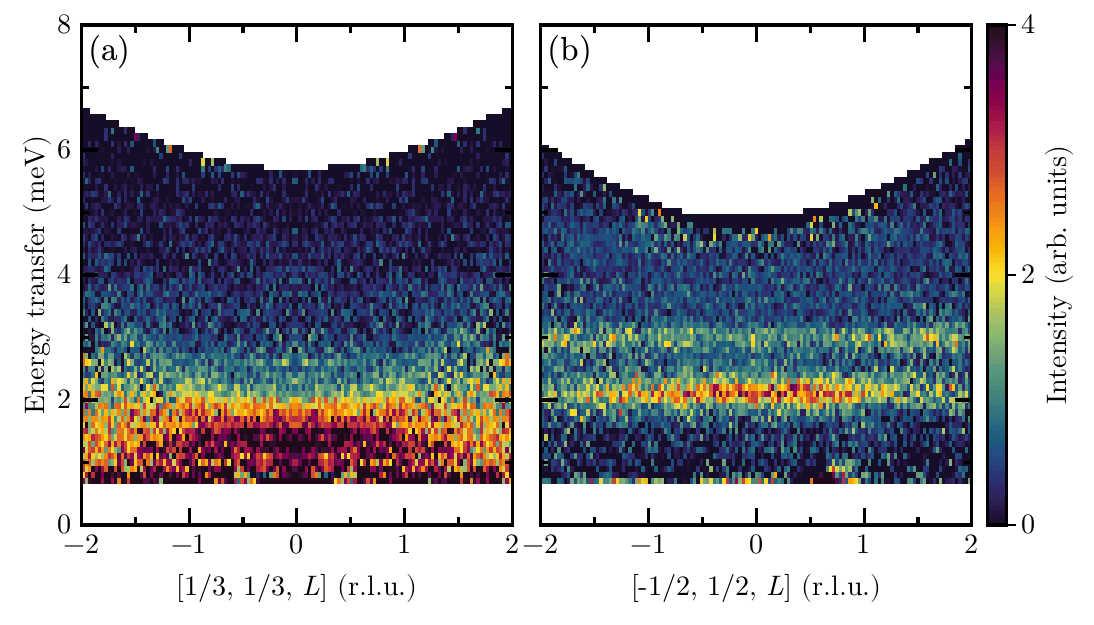}
        \caption{ Inelastic scattering intensity in the out-of-plane direction data with $E_i = 8$~meV at $T = 2$~K for two momentum point (a) ${\bf Q} = (1/3, 1/3, L)$ and (b) $(-1/2, 1/2, L)$, respectively. }
        \label{sfig:outofplane}
\end{figure}
\clearpage

\section{Spectrally-consistent non-linear spin-wave theory}
\label{sec:si:NLSWT}

Our implementation of spin-wave theory for the TLHAF follows Refs.~\cite{Chernyshev2006,Chernyshev2009}. These references were extended to calculate the DSSF in Ref.~\cite{Mourigal2013} and to the DSSF in the presence of XXZ anisotropy in Ref.~\cite{Maksimov2016}. We adhere to the supplementary material of Ref.~\cite{Maksimov2016} with $\Delta \neq 0$ and $H\!=\!0$. In this case, the harmonic magnon dispersion of the 2D TLHAF is obtained by diagonalizing the quadratic Hamiltonian, $\hat{\mathcal{H}}_2|{\bf q} \rangle=\varepsilon_{\bf q}|{\bf q} \rangle$. This yields,
\begin{eqnarray}
\varepsilon_{\bf q} & = & 3JS\,\sqrt{(1-\gamma_{\bf q})(1+2\Delta\gamma_{\bf q})} \\
\gamma_{\mathbf{q}} & = & \frac{1}{3} \left[ \cos q_x + 2\cos\!\left(\frac{q_x}{2}\right) \cos\!\left(\frac{\sqrt{3}q_y}{2}\right) \right].
\end{eqnarray} 
Here, ${\bf q} = (q_x,q_y)$ represents the hexagonal Brillouin zone. The harmonic dispersion includes corrections due to $1/S$ factors, which consist of a Hartree-Fock term ($\delta\varepsilon_{\bf q}^{\rm HF}$) and two self-energy corrections from source ($\Sigma_S(\bf q, \omega)$) and decay ($\Sigma_D(\bf q, \omega)$) of one-loop (one-bubble) magnon decay and source processes. These corrections are expressed as,
\begin{equation}
\Sigma ({\bf q},\omega) = \delta \varepsilon^{\rm HF}_{\bf q} + \Sigma_{D}({\bf q},\omega) + \Sigma_{S}({\bf q},\omega).
    \label{eq:genericse}
\end{equation}
The resulting dispersion curve is calculated self-consistently ``off-shell’’ as a solution of the Dyson equation, which yields,
\begin{equation}
    {\varepsilon}^\ast_{\bf q} = \varepsilon_{\bf q} + \delta \varepsilon^{\rm HF}_{\bf q} + \Sigma_{D}({\bf q},\varepsilon_{\bf q}^\ast) + \Sigma_{S}({\bf q}, \varepsilon_{\bf q}^\ast).\label{eq:offshellse}
\end{equation}

We observe that ${\varepsilon}^\ast_{\bf q}$ can acquire an imaginary part due to spontaneous decays, which introduces complexity in solving the Dyson equation \cite{Chernyshev2009}. However, given the small decay rate due to the values of $\Delta$ and $S$, additional care is not necessary in our case.

To obtain the DSSF corresponding to the renormalized single magnons, we first calculate its components in the local frame \cite{Mourigal2013} by introducing the one-magnon Green functions
\begin{eqnarray}
G({\bf q},\omega) & = & \frac{1}{\omega - \varepsilon_{\bf q} - \Sigma({\bf q},\omega) + i\delta}, \label{eq:greensidebands}
\end{eqnarray}
where $\delta$ denotes a small positive number. Here, $G$ encompasses the complete frequency-dependence of the magnon propagator in Eq.~\eqref{eq:genericse}, which we refer to as the “one-magnon sidebands.” We designate this framework as ${1/S}^\ast$-SWT, from which we compute the one-magnon spectral function as
\begin{equation}
{A}({\bf q},\omega) = -\frac{1}{\pi} {\rm Im} {G}({\bf q},\omega),
\label{eq:spectral}
\end{equation}
resulting in the transverse components of the DSSF in the local reference frame $\{x,y,z\}$ as
\begin{eqnarray}
{\mathcal{S}}^{{x}{x}}({\bf q},\omega) & = & \frac{\pi}{2} S \Lambda_+^2 (u_{\bf q} + v_{\bf q})^2 {A}({\bf q},\omega), \nonumber \\
{\mathcal{S}}^{{y}{y}}({\bf q},\omega) & = & \frac{\pi}{2} S \Lambda_-^2 (u_{\bf q} - v_{\bf q})^2 {A}({\bf q},\omega), \nonumber \\
{\mathcal{S}}^{{x}{y}}({\bf q},\omega) & = & i\frac{\pi}{2} S \Lambda_+ \Lambda_- {A}({\bf q},\omega), \label{eq:transverse}
\end{eqnarray}
where $\Lambda_\pm$ are intensity coefficients and $u_{\bf q}, v_{\bf q}$ are the Bogoliubov coefficients~\cite{Maksimov2016}. In the longitudinal component of the DSSF, the local frame is dominated by two-magnon excitations, which interact through a one-loop process in $\hat{\mathcal{H}}_4$. However, this process is of order $1/S^2$, so it lies beyond our calculation for the one-magnon dispersion. At order $1/S$, the longitudinal DSSF is given by the density of states of two non-interacting magnons. However, this approach has a drawback: the transverse and longitudinal response functions are calculated using magnon dispersions of different accuracies, which can become significant when comparing their features as functions of energy. This deficiency can be easily corrected by calculating the two-magnon density of states in the longitudinal component using the renormalized off-shell one-magnon dispersions
\begin{equation}
{\mathcal{S}}^{{z}{z}}({\bf q},\omega) = \frac{1}{2} \! \displaystyle{\sum_{{\bf p} \in {\rm BZ}}} \! (u_{\bf p} v_{\bf p-q} + v_{\bf p} u_{\bf p-q})^2
    \delta(\omega - {\varepsilon}^\ast_{\bf p} - {\varepsilon}^\ast_{\bf p-q}). \label{eq:longitudinal}
\end{equation}

 Our calculations were performed in Julia. All momentum-space integrals were implemented using h-adaptive multidimensional numerical integration using the {\scshape hcubature.jl}~\cite{HCUBATURE} package. Hartree-Fock energies were calculated with 7-digit precision, while self-energies and two-magnon calculations were calculated with 4-digit precision. We used and $\delta=0.06$~meV throughout.

\section{Determination of the goodness of fit with experimental data}
\label{sec:si:chi2}
The goodness of fit between $1/S^\ast$-SWT and the neutron scattering data was performed using the reduced $\chi^2$ loss function:
\begin{eqnarray}
    && \chi^2 = \frac{1}{N_{\rm fit}}\sum_{i,j}\frac{\left[{I}^{\rm dat}(q_i,\omega_j) - A\times{I}^{\rm sim}(q_i, \omega_j) \right]^2}{\sigma_{i,j}^2} \nonumber
\end{eqnarray}
where ${I}^{\rm dat}(q_i,\omega_j)$ is the inelastic neutron scattering intensity of the data for momentum index $i$, and energy index $j$; ${I}^{\rm sim}$ is the simulated inelastic neutron scattering intensity with the same indices; $\sigma_{i,j}$ is the error of the data; $A$ is overall scale factors between the simulations and the experimental data, which are uniquely determined by solving the linear equation: $\partial\chi^2/\partial{A} = 0$~\cite{Proffen1997}; and $N_{\rm fit}$ is the number of $(i,j)$ observations minus the number of free parameters, $N_{\rm fit} \approx N_{\rm obs}$.

\section{Beyond Mean-Field Schwinger Boson theory}
\label{sec:si:SBT}

The $S=1$ spin Hamiltonian of the  XXZ-model with nearest-neighbor exchange  interactions is 
\begin{equation}
    \mathcal{\hat H} = \sum_{\langle ij \rangle} J_{ij} (\hat S_{i}^x \hat S_{j}^x + \hat S_{i}^y \hat S_{j}^y + \Delta \hat S_{i}^z \hat S_{j}^z ).
\end{equation}
with $0<\Delta<1$ for easy-plane anisotropy. The spin operators are expressed in terms of Schwinger bosons,
\begin{equation}
    \hat {\bm S}_i = \frac{1}{2} \hat {\bm b}_{i}^\dag {\bm \sigma} \hat {\bm b}_{i}
\end{equation}
where $\hat {\bm b}^\dag = (\hat b_{i\uparrow}, \ \hat b_{i\downarrow})$ and $\bm \sigma \equiv (\sigma^x,\sigma^y,\sigma^z)$ is the vector of Pauli matrices. To project the Hilbert space described by the bosonic operators, $\hat b_{i\uparrow}$ and $\hat b_{i\downarrow}$, onto the physical space where the spin size is $S^2 = S (S + 1)$, the bosonic operators must satisfy the local constraint $\hat b_{i\uparrow}^\dag \hat b_{i\uparrow}^{} + \hat b_{i\downarrow}^\dag \hat b_{i\downarrow}^{} = 2S$.

In the Schwinger boson theory (SBT), the spin-spin interaction is expressed in terms of 
bond operators as \cite{Ghioldi2022,Scheie2023_KYbSe2_NatPhys,Zhang2022}
\begin{align}\label{eq:spin-spin}
    J_{ij} (\hat S_{i}^x \hat S_{j}^x + \hat S_{i}^y \hat S_{j}^y + \Delta \hat S_{i}^z \hat S_{j}^z ) 
    =  - J_{ij}^A \ \hat A_{ij}^{\dag} \hat A_{ij}^{} \ + J_{ij}^B  :\!\hat B_{ij}^{\dag} \hat B_{ij}^{}\!: - J_{ij}^C :\!\hat C_{ij}^{\dag} \hat C_{ij}^{}\!: + J_{ij}^D  \hat D_{ij}^{\dag} \hat D_{ij}^{}  + J_{ij}^S  S^2
\end{align}
with $J_{ij}^A = J_{ij}(1+\Delta)(1+ \varepsilon_1 )/2 $, $J_{ij}^B = J_{ij}(1+\Delta)(1- \varepsilon_1 )/2 $, $J_{ij}^C = J_{ij}(1-\Delta)(1+ \varepsilon_2 )/2 $, $J_{ij}^D = J_{ij}(1-\Delta)(1- \varepsilon_2 )/2 $, and $J_{ij}^S = J_{ij}[ (1+\Delta)\varepsilon_1 + (1-\Delta)\varepsilon_2 ]/2$.
The global SU(2)-invariant bond operators $ \hat A_{ij}^{\dag}$ and $ \hat B_{ij}^{\dag}$ are defined as
\begin{align}
    \hat A_{ij}^{\dag} = \frac{1}{2} ( \hat b_{i\uparrow}^{\dag} \hat b_{j \downarrow}^{\dag} - \hat b_{i\downarrow}^{\dag} \hat b_{j \uparrow}^{\dag} ) , \text{ and }
    & \hat B_{ij}^{\dag} = \frac{1}{2} (\hat b_{i\uparrow}^{\dag} \hat b_{j \uparrow}^{} + \hat b_{i\downarrow}^{\dag} \hat b_{j \downarrow}^{}  )
\end{align}
to account for the isotropic interaction. To account for the easy-plane anisotropy, the global U(1)-invariant operators
\begin{align}
    \hat C_{ij}^{\dag} = \frac{1}{2} ( \hat b_{i\uparrow}^{\dag} \hat b_{j \uparrow}^{} - \hat b_{i\downarrow}^{\dag} \hat b_{j \downarrow}^{} )  , \text{ and }
    & \hat D_{ij}^{\dag} = \frac{1}{2} ( \hat b_{i\uparrow}^{\dag} \hat b_{j \downarrow}^{\dag} + \hat b_{i\downarrow}^{\dag} \hat b_{j \uparrow}^{\dag} )
\end{align}
are introduced~\cite{Trumper1997}.

The bond-operator description introduces the continuous parameters $\varepsilon_1$ and $\varepsilon_2$, which parameterize distinct saddle-point (SP) approximations by modulating the relative weights of the anomalous pairing ($\hat A_{ij}^{\dag}$ and $\hat D_{ij}^{\dag}$) and normal hopping ($\hat B_{ij}^{\dag}$ and $\hat C_{ij}^{\dag}$) operators. As demonstrated in Ref.~\cite{Zhang2019}, these parameters primarily affect the two-spinon continuum, while the magnon dispersion exhibits significantly lower sensitivity to $\varepsilon_1$ and $\varepsilon_2$. This sensitivity further diminishes as the spin $S$ increases; indeed, in the large-$S$ limit, the SBT becomes $\varepsilon$-invariant and converges to LSWT. Generally, the spinon bandwidth increases with larger $\varepsilon_1$, while the magnon poles experience a slight drift. When the magnon poles emerge within the continuum, the peaks acquire a finite broadening; conversely, poles situated below the continuum remain well-defined delta-functions. The effect of $\varepsilon_2$ is primarily localized at the magnon gap produced by the anisotropy $\Delta$.

For the $S=1/2$ limit~\cite{Scheie2023_KYbSe2_NatPhys,Park2024BLCTO}, these parameters were previously determined by fitting the magnon dispersion to results obtained via series expansion methods~\cite{Ghioldi2015,Oitmaa2020,Zheng2006}. In the present $S=1$ case, we instead optimize the parameters ($\varepsilon_1=0.269$ and $\varepsilon_2=0$) to match the magnon bandwidth calculated within $1/S^\ast$-SWT. This fitting procedure subsequently defines both the magnon peak widths and the profile of the quasi-free two-spinon continuum (the free two-spinon continuum of the SP is slightly modified beyond the mean-field approximation).

To compute the dynamical spin structure factor (DSSF), we employ a large-$N$ theory ($N$ being the number of flavors in the boson fields, which is taken as $2$ for our physical system) based on the Feynman path integral formulation~\cite{Arovas1988,Auerbach1994}. In this formulation, the partition function is written as:
\begin{equation}
    \mathcal{Z}[h] = \int [D\bar b b][D\lambda] e^{-\int_{0}^{\beta} d\tau [\sum_{i\sigma} \bar b_{i\sigma} \partial_\tau b_{i\sigma} + \mathcal{H}(\bar b,b) + i \sum_{i} \lambda_i (\sum_{\sigma} \bar b_{i\sigma} b_{i\sigma} - 2S) + \sum_{\mu \alpha \beta i}  h_{i}^{\mu}(\tau) \frac{1}{2} \bar b_{i\alpha}^{} \sigma_{\alpha\beta}^{\mu} b_{i \beta} ] }
\end{equation}
where the complex fields $\bar b$ and $b$ are the eigenvalues of the Schwinger boson operators acting on coherent states, $\beta=1/k_BT$, $\lambda$ is the Lagrange multiplier introduced to impose the constraint of $2S$ bosons per site, and $\bm h$ is the external magnetic field coupled to the spins. Due to the quartic nature of the Hamiltonian in the boson fields, we use the Hubbard-Stratonovich (HS) transformation to decouple the bosonic fields,
\begin{equation}
    e^{J_{ij} \bar X_{ij} X_{ij} } = \int \frac{d\bar W_{ij}^{X} dW_{ij}^{X}}{2\pi i} \ e^{( -\bar W_{ij}^{X} W_{ij}^{X} +  \sqrt{|J_{ij}|} \bar X_{ij} W_{ij}^{X} + {\rm sgn}(J_{ij}) \sqrt{|J_{ij}|} \bar W_{ij}^{X} X_{ij} ) }
\end{equation}
where an auxiliary HS field $W^X$ is introduced for each bond field $X=A,B,C,D$. The phases of these HS fields and the Lagrange multiplier are the gauge fields of the theory. By integrating out the resulting quadratic action in the bosonic fields, we obtain the exact partition function 
\begin{equation}
    \mathcal{Z}[h] = \int [D\bar \phi D\phi] \ e^{-S_{\rm eff}(\bar \phi, \phi, h)} 
\end{equation}
where $S_{\rm eff}$ is the effective action which is a function of the auxiliary fields $\bar \phi_\alpha = (W_{ij}^X, \bar W_{ij}^X, \lambda) $.

The leading-order contribution in the large-$N$ expansion corresponds to the SP approximation, defined by the conditions:
\begin{equation}
    \frac{\partial S_{\rm eff}(\bar \phi, \phi, h=0)}{\partial \phi_\alpha} \bigg|_{\rm sp} = \frac{\partial S_{\rm eff}(\bar \phi, \phi, h=0)}{\partial \bar \phi_\alpha} \bigg|_{\rm sp} = 0.
\end{equation}
For the triangular lattice antiferromagnet studied here, long-range magnetic order emerges at the SP level via the Bose-Einstein condensation of Schwinger bosons. As established in Ref.~\cite{Zhang2022}, the selection of a non-fragmented condensate is essential to ensure a finite contribution from the $1/N$ corrections introduced below. 

The Dynamical Spin Susceptibility is computed from the partition function as
\begin{equation}
    \scalebox{1.2}{$\chi$}_{\mu \nu} (\mathbf{Q},\omega) = \frac{ \partial^2 \ln \mathcal{Z}[h] }{ \partial h_{\mathbf{Q},\omega}^{\mu} \ \partial h_{-\mathbf{Q}, -\omega}^{\nu} }
\end{equation}
where $h_{\mathbf{Q},\omega}^{\mu}$ is the $\mu$ component of the external magnetic field in momentum and frequency space.
Expanding the effective action as outlined in Ref.~\cite{Ghioldi2018}, we obtain the dynamical susceptibility beyond the SP approximation by incorporating the Gaussian fluctuations of the auxiliary fields,
\begin{equation}\label{eq:dss}
    \scalebox{1.2}{$\chi$}_{\mu \nu} (\mathbf{Q},\omega) = \scalebox{1.2}{$\chi$}_{\mu \nu}^{sp} (\mathbf{Q},\omega) + \scalebox{1.2}{$\chi$}_{\mu \nu}^{fl} (\mathbf{Q},\omega) 
\end{equation}
where ${\chi}_{\mu \nu}^{sp}$ is the SP contribution, and $\chi_{\mu \nu}^{fl}$ is a 1/$N$ correction. Notably, $\chi_{\mu \nu}^{fl}$ includes a contribution of order one that cancels the residues of the unphysical single-spinon poles in ${\chi}_{\mu \nu}^{sp}$, effectively acting as a counter-diagram to the SP result~\cite{Zhang2022}.

\begin{figure}[h!]
    \centering
    \includegraphics{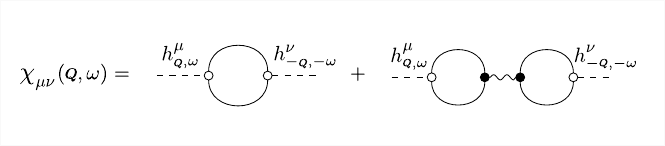}
    \caption{Diagrammatic expansion of the dynamical spin susceptibility [Eq.~(\ref{eq:dss})]. The first term on the right-hand side represents the SP contribution, while the second term captures the Gaussian fluctuations of the auxiliary fields. In the presence of a condensate, this second term acts as a counter-diagram to the SP term in addition to new poles. Dashed lines denote the external fields coupled via external vertices (open circles) $u^\mu$. Solid lines represent the free-spinon propagator $\mathcal{G}^{\rm sp}$ at the SP level, and solid circles denote internal vertices $v_{\phi}$. Wavy lines depict the propagator of the auxiliary fields (RPA propagator), whose poles correspond to the magnon excitations of the theory.}
    \label{fig:diag}
\end{figure}

These two contributions are diagrammatically represented in Fig. (\ref{fig:diag}). The solid lines denote the free-spinon propagator $\mathcal{G}^{\rm sp}$, the external vertices $u^{\mu}(\mathbf{Q}, \omega)\equiv \partial \mathcal{G}^{-1}/\partial h_{\mathbf{Q}, \omega}^\mu$ couple the spinons to the external fields $h_{\mathbf{Q},\omega}^{\mu}$. The internal vertices $v_{\phi}\equiv \partial \mathcal{G}^{-1}/\partial \phi$ couple the spinons to the auxiliary fields $\phi$. Finally, the auxiliary field propagator, equivalent to the Random Phase Approximation (RPA) propagator, is given by the inverse of the fluctuation matrix: $S^{(2)}_{\alpha\beta} \equiv \frac{1}{2} (\partial^2 S_{\rm eff} / \partial \phi_\alpha \partial \phi_\beta)|_{\rm sp} $.

Finally, the zero-temperature fluctuation-dissipation theorem relates the imaginary part of the dynamical susceptibility to the DSSF:
\begin{equation}
S^{\mu \nu}(\bf Q, \omega) = \frac{1}{\pi} \text{Im}[\chi_{\mu \nu} (\bf Q,\omega)],
\end{equation}
for $\omega>0$. Here, $S^{\mu \nu}(\bf Q, \omega)$ represents the spectral quantity directly observable via inelastic neutron scattering experiments.

\clearpage

\section{Comparison between Non-linear spin-wave theory and Schwinger Boson theory}
\label{sec:si:NLSWTvsSBT}

\begin{figure*}[htb!]
        \centering
        \includegraphics[width=1.0\linewidth]{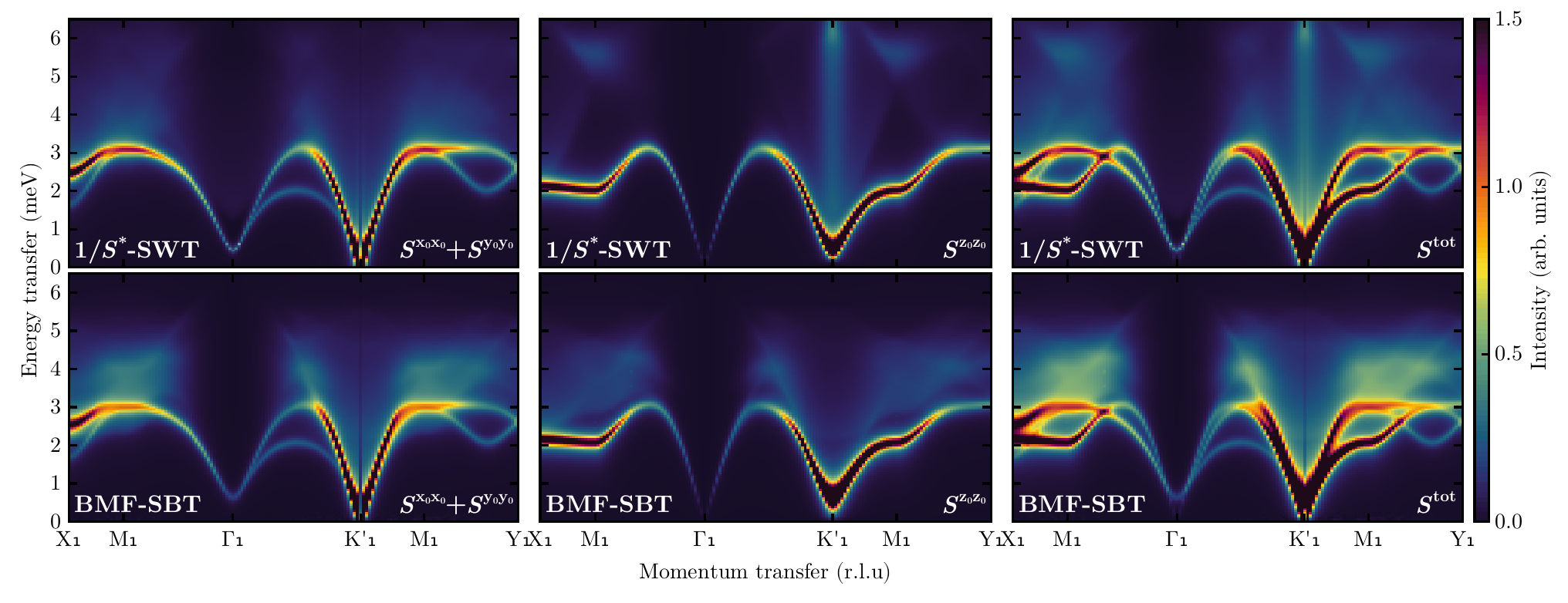}
        \caption{Comparison of the dynamical spin structure factor $\mathcal{S}^{\alpha\alpha}({\bf Q},\omega)$ ($\alpha = x_0, y_0, z_0$) between the $1/S^\ast$-SWT and BMF-SBT. $\mathcal{S}^{tot}$ indicates $\mathcal{S}^{x_0x_0} + \mathcal{S}^{y_0y_0} + \mathcal{S}^{z_0z_0}$. }
        \label{sfig:NLSWT_vs_SBT_1}
\end{figure*}

\begin{figure*}[htb!]
        \centering
        \includegraphics[width=1.0\linewidth]{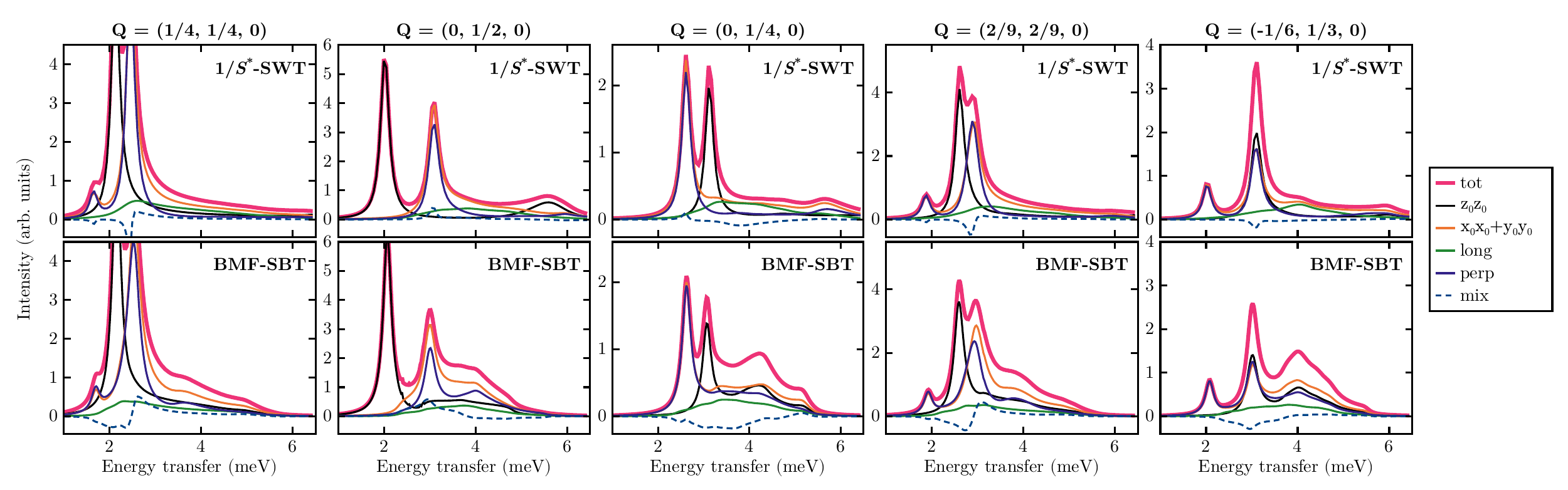}
        \caption{Channel-by-channel comparison of the dynamical spin structure factor $\mathcal{S}^{\alpha\alpha}({\bf Q},\omega)$ ($\alpha = x_0, y_0, z_0, x, y, z$) between the $1/S^\ast$-SWT and BMF-SBT at given momentum $\bf Q$. The 'long', 'perp', and 'mix' indicate the longitudinal $(xx + yy)$, transverse $(zz)$, and mixed $(xy)$ channels of the dynamical spin structure factor at the local frame $\{x,y,z\}$ defined in Eq.~\ref{eq:transverse} and Eq.~\ref{eq:longitudinal}, respectively.}
        \label{sfig:NLSWT_vs_SBT_2}
\end{figure*}

In this section, we compared the dynamical spin structure factor of two different approaches, $1/S^\ast$-SWT and BMF-SBT, without any integration effects or instrumental resolutions. Fig.~\ref{sfig:NLSWT_vs_SBT_1} shows the comparison of the dynamical spin structure factor $\mathcal{S}^{\alpha\alpha}({\bf Q},\omega)$ ($\alpha = x_0, y_0, z_0$) between the $1/S^\ast$-SWT and BMF-SBT at the lab frame $\{x_0, y_0, z_0\}$. Fig.~\ref{sfig:NLSWT_vs_SBT_2} presents the constant-momentum plots of channel-by-channel comparison for given momentum points \textbf{Q}.

\clearpage


\section{Landau-Lifshitz dynamics}
\label{sec:si:QLLD}

The finite-temperature response of our quantum model ($\mathcal{H}_{\rm Q} \equiv \mathcal{H}_{\rm min}$) was obtained in {\scshape Sunny.jl} using classical Landau-Lifshitz dynamics (LLD) \cite{Dahlbom2022, Sunny2025}. In these simulations the spin system is assumed to be a product state $|{\bf \Omega} \rangle = \otimes_i |{\bf \Omega}_i\rangle$ over SU(2) coherent states $|{\bf \Omega}_i\rangle$ representing dipolar spins operators $\hat{{\bf S}}_i$, which are replaced with classical vectors  ${\bf \Omega}_i = \langle  {\bf \Omega}_i | \hat{\bf S}_i |{\bf \Omega}_i \rangle$. The time-dependent dynamics at finite temperature are calculated using the stochastic Landau-Lifshitz-Gilbert (LLG) equation:
\begin{eqnarray}
    \frac{d\mathbf{\Omega}}{dt} = -\mathbf{\Omega}\times\left[\mathbf{\xi}(T)+\frac{dH}{d\mathbf{\Omega}}-\lambda\left(\mathbf{\Omega}\times\frac{dH}{d\mathbf{S}}\right)\right],
    \label{eq:stochastic_LLG_eq}
\end{eqnarray}
where $H =  \langle {\bf \Omega} | \mathcal{H}_Q | {\bf \Omega} \rangle$ in the large-$S$ limit, $\mathbf{\xi}(T)$ is a temperature-dependent Gaussian white noise, and $\lambda$ = 0.1 sets the coupling strength between the system and the thermal bath. 

The simulations were performed using a $60\times60\times1$ super-cell of the conventional chemical unit cell with periodic boundary conditions ($N=3,600$ spins) to calculate the neutron scattering response. The system was first thermalized by performing 5,000 Langevin time steps with a time step $\Delta t = 0.013$ meV$^{-1}$. After thermalization, 50 spin configurations were sampled using the LLG equation, with each configuration separated by 1,000 Langevin time steps to ensure decorrelation. 

To emulate the quantum dynamical spin structure factor (DSSF) from the classical spin trajectories calculated by LLD, we apply two temperature-dependent corrections~\cite{Kim2025Q2C, Dahlbom2024, Park2024SL, Kim2026RuCl3}. First, after performing the Fourier transform of the time trajectories of the real-space system, we apply the standard quantum-to-classical correspondence factor~\cite{Schofield1960,Zhang2019LLD},
\begin{eqnarray}
    \mathcal{S}_Q^{\mu\upsilon}(\mathbf{Q}, \omega) = \mathrm{sgn}(\omega)\frac{\hbar\omega}{k_BT}\frac{1}{1-e^{\hbar\omega/k_BT}}\mathcal{S}_{\rm cl}^{\mu\upsilon}(\mathbf{Q},\omega).
\end{eqnarray}
where $(\mu,\upsilon) \in (x_0,y_0,z_0)$, $\mathcal{S}_{\rm cl}$ is the classical DSSF from LLD, and $\mathcal{S}_{\rm Q}$ is the quantum-equivalent DSSF. Second, we rescale the spin length in the classical system, $|{\bf \Omega}_i|(T) = \kappa(T) S$, where the factor $\kappa(T)$ is chosen so that the DSSF satisfies the zeroth-order quantum sum rule at any given temperature $T$,
\begin{eqnarray}
    \sum_{\alpha}\int_{-\infty}^\infty d\omega\int d\mathbf{Q}\, \mathcal{S}_{\rm Q}^{\alpha\alpha}(\mathbf{Q},\omega;T) = NS(S+1).
\end{eqnarray}
Once $\kappa(T)$ is known from an initial simulation at a given target temperature $T_{\rm sim}$, the spin system is thermalized (in Langevin dynamics) and evolved (in Landau-Lifshitz dynamics) with the \textit{rescaled spin length}, and the quantum-to-classical correspondence is applied to calculate the $\kappa$-corrected quantum-equivalent DSSF. This spin-rescaling process effectively introduces a renormalization between the bath temperature ($T_{\rm sim}$) and the effective temperature experienced by the spin system during thermalization and dynamical evolution. As this difference increases with decreasing $T_{\rm sim}$, we empirically associate this effect with a correction between Boltzmann and Bose-Einstein thermal statistics. 

Throughout the manuscript, the neutron scattering intensity was computed from the DSSF using:
\begin{eqnarray}
    {I}({\bf Q},\omega) \propto f^2(|{\bf Q}|)\sum_{\mu, \upsilon}\left[\delta_{\mu\upsilon} - \frac{Q_\mu Q_\upsilon}{|{\bf Q}|^2}\right]\mathcal{S}^{\mu \upsilon}(\mathbf{Q}, \omega)
\end{eqnarray}
where $f(|{\bf Q}|)$ is the magnetic form factor, $Q^\alpha$ is the projection of the momentum onto the spin components in the local cubic coordinate system also used for the spin Hamtilonian, and $\mathcal{S}^{\mu\upsilon}(\mathbf{Q},\omega)$ is the computed dynamical spin structure factor at momentum \textbf{q} and energy $\omega$. The neutron scattering intensity was integrated along the out-of-plane direction by assuming $\mathcal{S}^{\mu\upsilon}(\mathbf{Q},\omega)$ is constant along the out-of-plane.

\begin{figure*}[htb!]
        \centering
        \includegraphics[width=0.5\linewidth]{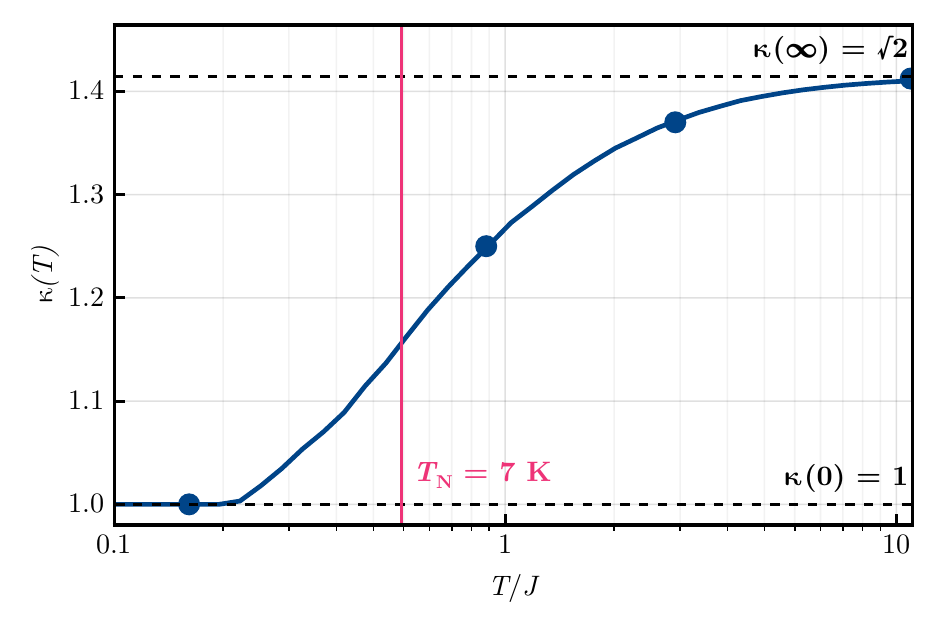}
        \caption{Temperature dependence of $\kappa(T)$ rescaling factor by satisfying the quantum sum-rule. Blue dots present the value of $\kappa(T)$ used in the DSSF calculation for given temperatures. Black dashed lines are the two limits of $\kappa(T)$ at zero temperature and infinite temperature. The magenta vertical line indicates the Neel temperature $T_N$ = 7 K of the \knso.}
        \label{sfig:kappa_T_dep}
\end{figure*}
\begin{figure*}[htb!]
        \centering
        \includegraphics[width=1.0\linewidth]{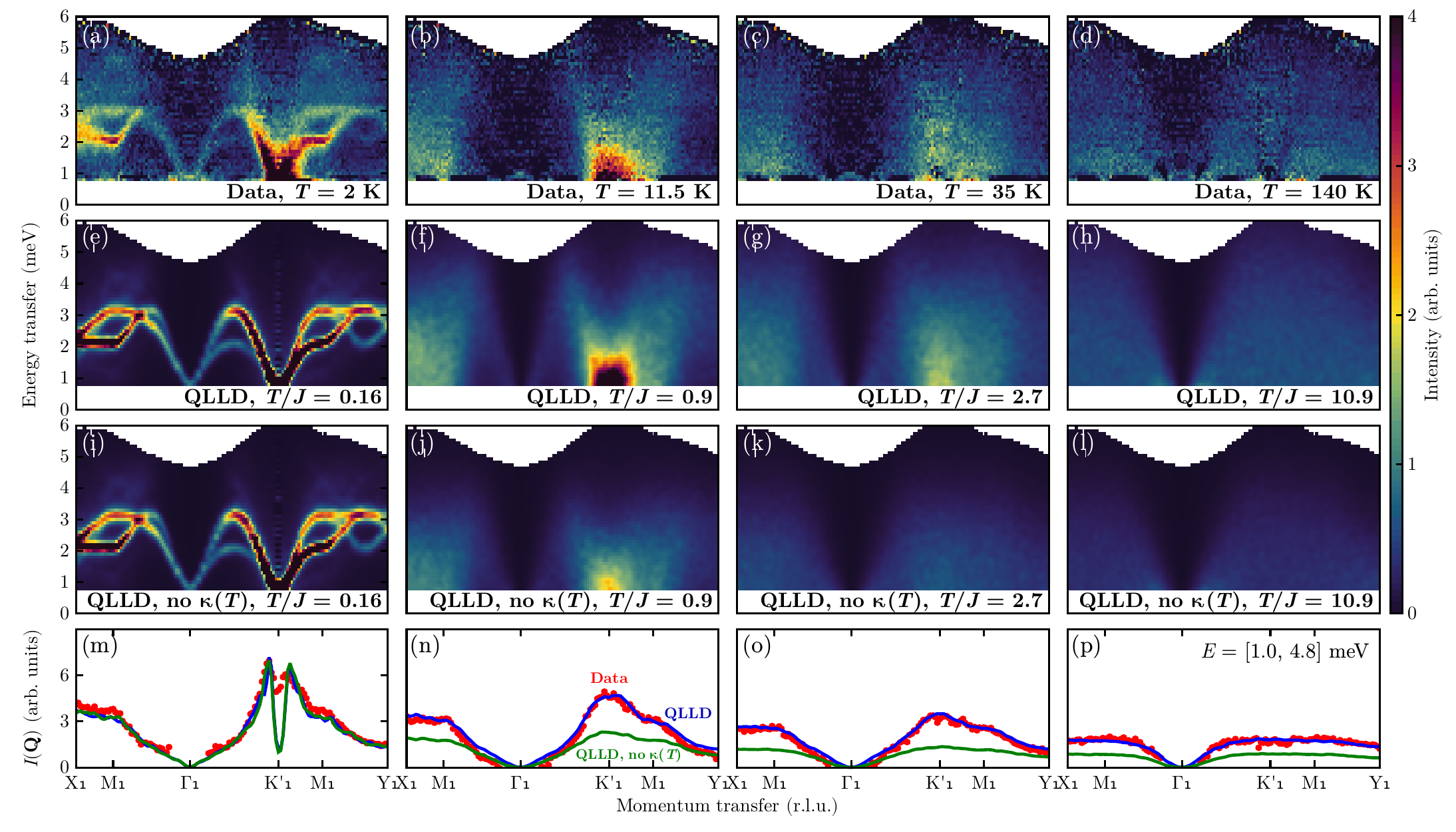}
        \caption{Full temperature dependence of the spin dynamics of \knso (a-d) and its comparison with QLLD simulations with (e-h) and without (i-l) quantum sum-rule correction (i.e. $\kappa(T)$ rescaling). Bottom row (m-p) shows the energy-integrated intensity with $E$ = [1.0, 4.8] meV. Red dots present the experimental data, blue and green solid lines are energy-integrated QLLD with and without $\kappa(T)$ rescaling, respectively.}
        \label{sfig:LLD_T_dep_full}
\end{figure*}

\end{document}